\documentclass{PoS}

\usepackage{amsmath}

\title{Majorons as cold light dark matter}

\ShortTitle{Majorons as cold light dark matter}

\author{\speaker{Julian Heeck}\\
        Service de Physique Th\'eorique, Universit\'e Libre de Bruxelles, CP225, 1050 Brussels, Belgium\\
        E-mail: \email{julian.heeck@ulb.ac.be}}

\abstract{Majorons are the Goldstone bosons of spontaneously broken lepton number and hence intimately connected to Majorana neutrino masses. Since all majoron couplings are heavily suppressed by the seesaw scale they are interesting candidates for long-lived dark matter. The signature decay into two mono-energetic neutrinos is potentially detectable with neutrino detectors for majoron masses above MeV and complementary to the loop-induced decays into visible particles. The mass range between keV and MeV can only be probed indirectly with the majoron decay into two photons; keV-scale majorons can be warm or cold dark matter depending on the underlying freeze-in mechanism.
	}

\FullConference{Neutrino Oscillation Workshop (NOW2018)\\
		9 - 16 September, 2018\\
		Rosa Marina (Ostuni, Brindisi, Italy)}

\begin{document}

\section{Introduction}

The accidental global lepton-number symmetry $U(1)_L$ could be broken \emph{spontaneously} rather than explicitly in the seesaw mechanism of Majorana neutrino masses, in which case one predicts a massless Goldstone boson $J$ with couplings suppressed by the $U(1)_L$ breaking scale $f$~\cite{Chikashige:1980ui,Schechter:1981cv}. Assuming the full Lagrangian (including gravity) contains some small explicit $U(1)_L$-breaking terms, this majoron $J$ becomes a \emph{pseudo}-Goldstone boson with mass $m_J$, taken as a free parameter in the following. As realized already long ago~\cite{Rothstein:1992rh,Berezinsky:1993fm}, the heavily suppressed couplings $\propto 1/f$ make the majoron potentially long-lived enough to act as dark matter (DM), with signature tree-level decay channel $J\to \nu\nu$ and loop-level decay channels into visible particles. The signatures of this DM candidate depend on whether $m_J$ is below or above MeV, as discussed in the following.

\section{Majoron with mass above MeV}

Assuming the correct DM density to be produced via a freeze-in mechanism~\cite{Frigerio:2011in}, we can discuss the signatures of unstable majoron DM with mass above MeV~\cite{Garcia-Cely:2017oco,Heeck:2017wgr,Heeck:2017kxw}.
The most conservative constraint on $\Gamma (J\to \nu\nu)$ comes from cosmology and requires the DM lifetime to be $\mathcal{O}(10)$ times the age of our Universe~\cite{Lattanzi:2007ux,Audren:2014bca,Poulin:2016nat}. However, for DM masses above $\sim 4$\,MeV one can actually search for the mono-energetic neutrinos with neutrino detectors such as Borexino, KamLAND, and Super-Kamiokande, which can give far better constraints and warrant dedicated experimental analyses~\cite{PalomaresRuiz:2007ry,Garcia-Cely:2017oco}. With upcoming experiments such as Hyper-Kamiokande, JUNO, and DUNE, prospects for improvement in this direction are very good and could make it possible to detect majoron DM via the smoking gun neutrino lines.

Since neutrinos are part of an $SU(2)_L$ doublet, it is ill-advised to discuss neutrino couplings and not charged leptons. In the simplest majoron model, the $J$ couplings to charged leptons and quarks arise at the one-loop level~\cite{Chikashige:1980ui,Pilaftsis:1993af}, while that to photons at two-loop~\cite{Garcia-Cely:2017oco}. The corresponding decays $J\to \overline{\ell}\ell', \overline{q}q, \gamma\gamma$ are strongly constrained by their impact on the cosmic microwave background~\cite{Slatyer:2016qyl} as well as by $\gamma$-ray telescopes such as Fermi-LAT.
Interestingly, these visible decay channels depend on different model parameters than the tree-level majoron decay into neutrinos, making it impossible to compare these constraints~\cite{Garcia-Cely:2017oco}. In other words, the neutrino channel is \emph{complementary} to the usual visible indirect-detection signatures and could in fact be the discovery channel of majoron DM!

\section{Majoron with mass below MeV}

For sub-MeV majoron masses, only the decay $J\to\gamma\gamma$ remains as a promising indirect detection signature~\cite{Berezinsky:1993fm,Bazzocchi:2008fh,Lattanzi:2013uza}, seeing as $J\to \nu\nu$ would give neutrinos of too-low energy to induce inverse beta decay, thus making line-searches challenging. For DM masses below $\mathcal{O}(10)$\,keV one can have an additional effect on structure formation because DM could be \emph{warm}~\cite{Kuo:2018fgw}. This is precisely the region of interest for the tantalizing line at $E_\gamma \simeq 3.55$\,keV observed in Refs.~\cite{Bulbul:2014sua,Boyarsky:2014ska}, which would hint at a $7$\,keV majoron. The significance and origin of this line are rather controversial, see e.g.~Refs.~\cite{Adhikari:2016bei,Abazajian:2017tcc}, but it still serves as a good benchmark value.

For such keV-scale DM masses one has to be careful not to violate structure-formation constraints from the Lyman-$\alpha$ forest, which effectively put limits on the DM free-streaming length. This depends on the DM \emph{production} mechanism and is thus model dependent, with many models being incompatible with a 7\,keV DM mass~\cite{Merle:2014xpa}. As shown in Ref.~\cite{Heeck:2017xbu}, there however exist production mechanisms that can produce keV DM in a very cold way, thus allowing to accommodate $\gamma$-lines down to keV without affecting structure formation. These mechanisms are necessarily of the freeze-in type and require a calculation of the DM momentum distribution in order to assess the mean DM momentum, which is the appropriate quantity that governs the DM warmness~\cite{Heeck:2017xbu,Boulebnane:2017fxw}. As one simple example for such a mechanism let us consider the DM production via the decay $A\to B \, \text{DM}$, where $A$ and $B$ are heavy particles in thermal equilibrium and the DM candidate is much lighter than $m_A-m_B$. In the $A$ rest frame it is obvious that the DM momentum becomes smaller the more degenerate $A$ and $B$ are, simply due to phase-space suppression. It turns out that this feature survives in the thermal bath, leading to a mean DM momentum $\langle p/T\rangle \simeq \tfrac{5}{2} \, (1-m_B^2/m_A^2)$ at the time of DM production. Even a mild degeneracy of $A$ and $B$ can thus suppress $\langle p/T\rangle$ far below the typical $\mathcal{O}(3)$ values given by other mechanisms. This implies a very short free-streaming length which evades any structure-formation constraints even for keV DM mass, while still allowing for $J\to\gamma\gamma$ signatures. The required setup, a light DM particle with off-diagonal coupling to two quasi-degenerate heavy particles, can naturally be found in inverse-seesaw majoron models, discussed in detail in Ref.~\cite{Boulebnane:2017fxw}, where $A$ and $B$ form a sterile neutrino pseudo-Dirac pair. 

\section{Conclusion}

Majorons make for interesting unstable DM candidates, intimately linked to the Majorana neutrino mass generation. The signature majoron decay into mono-energetic neutrinos is potentially detectable for energies above MeV and motivates dedicated searches with neutrino detectors, which are nicely complementary to searches for gamma-ray lines etc.
The parameter space \emph{below} MeV offers fewer testable channels, but is still constrained by x-ray lines and structure formation, e.g.~in the form of Lyman-$\alpha$ forest data. These two signatures can be decoupled depending on the DM production mechanism, making it in particular possible to have \emph{cold} keV DM, naturally occurring in majoronic inverse seesaw mechanisms.

\acknowledgments

I thank the NOW organizers for a smooth, interesting and stimulating conference.
This work was supported by the F.R.S.-FNRS.

\bibliographystyle{JHEP}
\bibliography{BIB}

\end{document}